\documentclass[11pt]{article}
\usepackage{amssymb,amsmath,graphicx}

\def\[{\begin{equation}}
\def\]{\end{equation}}
\def\d{{\rm d}}
\def\e{{\rm e}}
\let\~=\tilde
\def\sech{\mathop{\rm sech}\nolimits}
\def\csch{\mathop{\rm csch}\nolimits}
\def\Complex{{\mathbb{C}}}

\makeatletter \numberwithin{equation}{section}

\begin{document}

\title{Solitary waves and their linear stability in nonlinear lattices}
\author{By \emph{Guenbo Hwang, T.~R.~Akylas, and Jianke Yang}\footnote{
Address for correspondence: Prof. J. Yang, Department of Mathematics
and Statistics, University of Vermont, Burlington, VT 05401, USA;
e-mail: jyang@cems.uvm.edu}}
\date{ }
\maketitle

Solitary waves in a general nonlinear lattice are discussed,
employing as a model the nonlinear Schr\"odinger equation with a
spatially periodic nonlinear coefficient.  An asymptotic theory is
developed for long solitary waves, that span a large number of
lattice periods.  In this limit, the allowed positions of solitary
waves relative to the lattice, as well as their linear stability
properties, hinge upon a certain recurrence relation which contains
information beyond all orders of the usual two-scale perturbation
expansion.  It follows that only two such positions are permissible,
and of those two solitary waves, one is linearly stable and the
other unstable.  For a cosine lattice, in particular, the two
possible solitary waves are centered at a maximum or minimum of the
lattice, with the former being stable, and the analytical
predictions for the associated linear stability eigenvalues are in
excellent agreement with numerical results.  Furthermore, a
countable set of multi-solitary-wave bound states are constructed
analytically.  In spite of rather different physical settings, the
exponential asymptotics approach followed here is strikingly similar
to that taken in earlier studies of solitary wavepackets involving a
periodic carrier and a slowly-varying envelope, which underscores
the general value of this procedure for treating multi-scale
solitary-wave problems.

\bigskip

%%%%%%%%%%%%%%%%%%%%%%%%%%%%%%%%%%%%%%%%%%%%%%%%%%%%%%%%%%%%%%%%%%%%%%%%%%%%%%%
\section{Introduction}
%%%%%%%%%%%%%%%%%%%%%%%%%%%%%%%%%%%%%%%%%%%%%%%%%%%%%%%%%%%%%%%%%%%%%%%%%%%%%%%

The nonlinear Schr\"odinger (NLS) equation is fundamental to wave
propagation in fluid flows, optics, Bose--Einstein condensation
(BEC) and applied mathematics
\cite{Benney_Newell,Craik_FluidFlows,Ablowitz_book,BEC_GP,Kivshar_book,Yang_SIAM}.
When a linear periodic potential (so-called linear lattice) is
added, the resulting equation then models nonlinear beam
transmission in an optical medium with a periodic transverse
variation in the linear refractive index, and atom--atom interaction
in Bose--Einstein condensates loaded in an optical lattice
\cite{Christodoulides_review,BEC_review,Panos_book,Yang_Cambridge}.
Another interesting possibility, instead of a linear lattice, is to
allow the nonlinear coefficient of the NLS equation to vary
periodically in space. In optics, such a periodic nonlinear
coefficient (also called nonlinear lattice) would arise in the
propagation of laser beams in a medium whose nonlinear refractive
index is modulated in the transverse direction. The same model
equation also applies to the study of Bose--Einstein condensates in
a medium with a spatially-dependent scattering length. In optical
and BEC experiments, a linear lattice can be created by beam
interference \cite{BEC_review,Fleischer}. A nonlinear lattice in
optics can be created by femtosecond-laser writing in fused silica
\cite{Szameit}, and  in BEC such a lattice can be created by an
optical Feshbach-resonance modulation of the scattering length
\cite{BEC_Feshbach1,BEC_Feshbach2,BEC_nonlinear_lattice_experiment}.
Wave phenomena in linear lattices have been extensively studied in
the past decade (see \cite{Yang_Cambridge} for a review). In this
paper, our interest centers on wave phenomena in nonlinear lattices.

Wave phenomena in symmetric nonlinear lattices have been
investigated in
\cite{Malomed_nonlinear_lattice,Fibich_PhysicaD,Bishop_nonlinear_lattice}.
It was found that there exist solitary waves which are centered at a
local maximum or minimum of the nonlinear lattice. Short solitary
waves were found to be
stable/unstable when centered at a local maximum/minimum of the lattice
\cite{Fibich_PhysicaD,Bishop_nonlinear_lattice}, but the stability
analysis for long solitary waves, that extend over a large number of lattice periods, was not conclusive
\cite{Fibich_PhysicaD}. Bound states comprising several fundamental
solitary waves have also been reported numerically
\cite{Malomed_nonlinear_lattice}. In addition to these studies in
nonlinear lattices, work has also been done in the presence of both
linear and nonlinear lattices
\cite{Kevrekidis_lin_non_lattice,Perez_Garcia_lin_non_lattice,Pelinovsky_lin_non_lattice}
and in higher dimensions \cite{Fibich_PRL}.

In this paper, we make an analytical study of long solitary waves
and their linear stability properties in a general nonlinear
lattice. Recognizing that the coupling between a long solitary wave
and the relatively short nonlinear lattice is an exponentially small
effect, beyond all orders of the usual multiple-scale perturbation
expansion in powers of the long-wave parameter, our analysis makes
use of an exponential asymptotics technique. We show that long
solitary waves can only be located at two positions in one period of
the nonlinear lattice, regardless of the number of local maxima and
minima in it. If the lattice is symmetric, the solitary-wave
positions are simply the point of symmetry and half a lattice-period
away from it; but for a general asymmetric lattice, these positions
need to be determined by solving a certain recurrence relation that
contains information beyond all orders of the long-wave parameter.
Linear stability analysis of long solitary waves is also performed:
it follows from the same recurrence relation that one of the two
solitary waves is linearly stable, while the other is unstable. For
cosine lattices, in particular, the solitary wave centered at the
maximum/minimum of the lattice is linearly stable/unstable,
consistent with previous results
\cite{Fibich_PhysicaD,Bishop_nonlinear_lattice}. An analytical
formula for the linear-stability eigenvalues is also derived; as
expected, these eigenvalues are exponentially small in the long-wave
parameter.  The analytical predictions for the stability eigenvalues
are compared with direct numerical results and excellent agreement
is obtained. Lastly, we also show that an infinite number of
multi-solitary-wave stationary bound states exist in the nonlinear
lattice, and their analytical construction in terms of nonlocal
solitary waves is presented.

The exponential asymptotics procedure adopted in this paper closely
resembles that used in recent studies \cite{Hwang_stability,Akylas_bound_states}
 of gap solitons in a linear lattice (see also
\cite{Akylas_JFM1997,Calvo_stability} for the first application of
this technique to solitary wavepackets of the fifth-order KdV
equation). The common thread in linear and nonlinear lattices is
that the solitary wave is much longer than the period of the
underlying lattice; hence, the coupling between these different
scales is expected to be exponentially small, which invites similar
exponential asymptotics treatment. As a result, the behavior of
solitary waves in these rather different physical settings is
remarkably similar.

%%%%%%%%%%%%%%%%%%%%%%%%%%%%%%%%%%%%%%%%%%%%%%%%%%%%%%%%%%%%%%%%%%%%%%%%%%%%%%%
\section{Preliminaries}  \label{sec:preliminaries}
%%%%%%%%%%%%%%%%%%%%%%%%%%%%%%%%%%%%%%%%%%%%%%%%%%%%%%%%%%%%%%%%%%%%%%%%%%%%%%%
We study the NLS equation with a nonlinear lattice
\[
i\Psi_t+\Psi_{zz}+(1+g(z/\epsilon))|\Psi|^2\Psi=0\,,
\label{e:NLSE}
\]
where $g(z/\epsilon)$ is a periodic function which describes the
spatial variation of the nonlinear Kerr coefficient, and the
parameter $\epsilon>0$ controls the length scale of this  variation.
Throughout this article, $g(z/\epsilon)$ will be referred to as the
nonlinear lattice. Eq. (\ref{e:NLSE}) is a model for the spatial
propagation  of a laser beam in a medium whose nonlinear refractive
index is modulated periodically in the transverse direction (in this
context, $t$ is the direction of propagation), and for the dynamics
of Bose--Einstein condensates whose scattering length (the
counterpart of the nonlinear coefficient in (\ref{e:NLSE})) changes
periodically over space
\cite{BEC_GP,BEC_nonlinear_lattice_experiment}. Solitary-wave
solutions in Eq. (\ref{e:NLSE}) and their linear-stability
properties for even functions of $g(z/\epsilon)$ were investigated
in \cite{Fibich_PhysicaD,Bishop_nonlinear_lattice}. In the long-wave
limit $(\epsilon \ll 1)$, profiles of solitary waves that span many
lattice periods were determined by a multiscale perturbation
analysis \cite{Fibich_PhysicaD}, but their stability was not
ascertained.

In this article, we analytically study long solitary waves
($0<\epsilon \ll 1$) and their stability properties for general
forms of the nonlinear lattice $g(z/\epsilon)$. Specifically, denoting $z/\epsilon=x$, $g(x)$ is assumed to be periodic with
period $d$,
\[
g(x+d)=g(x)
\]
for all real values of $x$.  Also, without loss of generality, $g(x)$ is taken to have zero $x$-average:
\[
\langle g \rangle  \equiv \frac{1}{d} \int_0^d g(x) \d x=0\,.
\label{def_average}
\]

Solitary-wave solutions of Eq.~\eqref{e:NLSE} are sought in the form
\[
\Psi(z,t)=\psi(z)\e^{i\mu t}\,,
\]
where $\mu>0$ is the propagation constant, and the real-valued
function $\psi(z)$ solves the  ordinary differential
equation
\[
\psi_{zz}+(1+g(z/\epsilon))\psi^3-\mu\psi=0\,,\label{e:NLSEamp}
\]
under the boundary condition of $\psi\to 0$ as $|z|\to \infty$.

Here, we will fix $\mu=O(1)$, and determine the solitary wave
$\psi(z)$ as well as its linear stability for $0<\epsilon \ll 1$. In
this regime, the nonlinear lattice $g(z/\epsilon)$ is rapidly
varying and the wave profile $\psi(z)$, whose width is $O(1)$ for
$\mu=O(1)$, spans many lattice sites.  As $\epsilon\to 0$,
$g(z/\epsilon)$ features extremely rapid oscillations with zero mean
and the effect of the nonlinear lattice on the solitary wave
$\psi(z)$  can be dropped; thus, in this limit, $\psi(z)$ is
expected to approach the familiar solitary-wave solution of the
lattice-free NLS equation
\[
\psi(z) \to  a\sech{z-z_0\over \beta}\,,  \label{e:psilimit}
\]
where
\[
a=\sqrt{2\mu}, \quad  \beta=1/\sqrt{\mu},    \label{e:abeta}
\]
and $z_0$ denotes the location of the peak of the solitary wave
$\psi(z)$. When $0\ne \epsilon\ll 1$, the nonlinear lattice
$g(z/\epsilon)$ will have a weak but non-negligible effect, and
solitary waves bifurcate out from the limiting wave
(\ref{e:psilimit}); this bifurcation will be the main focus of our
investigation. It will be shown that such a bifurcation is possible
only when $z_0$ takes two special values relative to the nonlinear
lattice, resulting in two solitary-waves, out of which one is stable
and the other unstable.

It is worth mentioning that a different but equivalent analysis is
to  introduce scaled variables
\[
\widehat{\psi}=\epsilon \psi, \quad x=z/\epsilon, \quad
\widehat{\mu}=\epsilon^2 \mu, \label{e:2-8}
\]
so that Eq. (\ref{e:NLSEamp}) transforms into
\[
\widehat{\psi}_{xx}+(1+g(x)) \hspace{0.04cm}
\widehat{\psi}^3-\widehat{\mu}\widehat{\psi}=0\,. \label{e:NLSEampb}
\]
Thus, fixing $\mu=O(1)$ and varying $\epsilon\ll 1$ in our treatment
above corresponds to  fixing the lattice $g(x)$ and varying
$\widehat{\mu}\ll 1$ in Eq. (\ref{e:NLSEampb}). In this alternative
treatment, the nonlinear lattice $g(x)$ is no longer rapidly
varying, but for $\widehat{\mu}\ll 1$, solitary waves
$\widehat{\psi}(x)$ of Eq. (\ref{e:NLSEampb}) that bifurcate out
from the edge $\widehat{\mu}_0=0$ of the continuous spectrum (whose
linear eigenmode is a constant) have low amplitude and are long
compared to the lattice period. This type of solitary-wave
bifurcation is akin to that studied in \cite{Hwang_stability} with a
linear lattice.  In both treatments, however, the solitary wave is
of long extent and spans many lattice sites, hence the fundamental
feature of the problem remains the same.

\section{Multiscale perturbation solution}

We begin by reviewing the multiple-scale procedure followed
in~\cite{Fibich_PhysicaD} for solving equation~\eqref{e:NLSEamp}.
The solution $\psi$ to this equation contains two scales, reflected
in the `slow' variable $z$  and `fast' variable $x=z/\epsilon$.
Introducing explicitly these variables by writing $\psi=\psi(x,z)$,
Eq.~\eqref{e:NLSEamp} then becomes
\[
\bigg( \partial^2_z +\frac2\epsilon\partial_x\partial_z
+\frac1{\epsilon^2}\partial^2_x\bigg)\,\psi + (1+g(x))\psi^3-\mu\psi=0\,.
\label{e:eqmulitscale}
\]
Now we expand $\psi(x,z)$ into the two-scale perturbation series,
\[
\psi(x,z)  =\psi_0(x,z)+\epsilon \psi_1(x,z)+\epsilon^2 \psi_2(x,z)+\cdots\,.\label{e:psiexpansion}
\]
Substituting this expansion into Eq.~\eqref{e:eqmulitscale} and from
various orders of $\epsilon$, we get the following hierarchy of equations,
\begin{subequations}
\begin{align}
-\partial^2_x \psi_0 & =0\,,  \label{e:perteqa}\\
-\partial^2_x \psi_1 & =2\partial_x\partial_z \psi_0\,, \label{e:perteqb}\\
 -\partial^2_x \psi_2
&=\partial^2_z\psi_0+2\partial_x\partial_z \psi_1
+(1+g(x))\psi^3_0-\mu\psi_0 \,,\label{e:perteqc}\\
-\partial^2_x \psi_3 &=\partial^2_z\psi_1+2\partial_x\partial_z \psi_2
+3(1+g(x))\psi^2_0\psi_1-\mu\psi_1 \,,\label{e:perteqd}\\
-\partial^2_x \psi_4 &=\partial^2_z\psi_2+2\partial_x\partial_z \psi_3
+3(1+g(x))\psi^2_0\psi_2+3(1+g(x))\psi_0\psi^2_1-\mu\psi_2\,.\label{e:perteqe}
\end{align}
\end{subequations}
From Eq. (\ref{e:perteqa}) and the requirement that $\psi_0$ be
bounded, we get
\[
\psi_0=\~\psi_0(z).     \label{e:psi0}
\]
Substituting this equation into (\ref{e:perteqb}) and requiring
$\psi_1$ to be bounded, we get
\[
\psi_1=\~\psi_1(z).     \label{e:psi1}
\]
Equations (\ref{e:perteqc})--(\ref{e:perteqe}) are forced linear
equations and can be written in the unified form
\[
-\partial^2_x \psi_n=Q_n(x,z)\,,
\label{e:forcedeq}
\]
where $Q_n$ depends on $\psi_j$ with $j<n$. The $x$-dependence of
$Q_n$ derives from the nonlinear lattice $g(x)$ which is
$d$-periodic. Thus $Q_n$ is $d$-periodic in $x$ and can be expanded
into a Fourier series,
\[
Q_n(x,z)=\sum_{m=-\infty}^{\infty} q_m(z) \e^{2\pi i mx/d}.
\label{QFourier}
\]
We require solutions $\psi_n$ to be also $d$-periodic in $x$. The
necessary and sufficient condition for the existence of such a
solution is that the constant Fourier mode $q_0$ in (\ref{QFourier})
vanish,
\[   \label{e:q00}
q_0 = \langle Q_n \rangle =0\,.
\]
Here $\langle \cdot \rangle$ is the average with respect to $x$ as
defined in (\ref{def_average}). Then the solution $\psi_n$ to
Eq.~\eqref{e:forcedeq} is
\[
\psi_n=-\partial^{-2}_x Q_n(x,z)+\~\psi_n(z)\,,
\label{e:gensolnpsin}
\]
where
\[
\partial^{-2}_x Q_n=\sum_{m\ne 0} \bigg({2\pi i m\over d} \bigg)^{-2}
q_m(z) \e^{2\pi i mx/d}\,,     \label{e:dx2Qn}
\]
and $\~\psi_n$ is a function of $z$ which is determined by the
solvability condition (\ref{e:q00}) for the equation governing
$\psi_{n+2}$ (see below).

Next we determine $\~\psi_0(z)$ and $\~\psi_1(z)$. Specifically, to obtain $\~\psi_0(z)$,
we return to equation
(\ref{e:perteqc}). Substituting (\ref{e:psi1}) into
(\ref{e:perteqc}) and recalling that $g(x)$ has zero average (see
(\ref{def_average})), the solvability condition of this
equation gives
\[
\partial^2_z \~\psi_0+\~\psi_0^3-\mu \~\psi_0=0\,,    \label{e:psi0eq}
\]
whose solution is
\[
\~\psi_0=A(z) \equiv   a\sech{z-z_0\over \beta}, \label{e:envelopeA}
\]
where $a$ and $\beta$ are defined in (\ref{e:abeta}), and $z_0$ is a
constant. This result agrees with
(\ref{e:psilimit}), as anticipated earlier.

To determine the solution $\~\psi_1(z)$, we consider equation (\ref{e:perteqd}). Using the above expression for
$\~\psi_0$ and the zero average of $g(x)$, the solvability
condition for (\ref{e:perteqd}) gives
\[
\left({\d^2/\d z^2}-\mu+3A^2 \right) \~\psi_1 = 0\,,
\label{e:linzeq}
\]
whose solution is
\[
\~\psi_1=\zeta A'(z),
\]
where $\zeta$ is a constant. It is clear that $\~\psi_1$  simply
shifts the location of the center of the leading-order term,
$\~\psi_0=A(z)$, in the perturbation
expansion~\eqref{e:psiexpansion}. To remove the ambiguity in the
position of $\~\psi_0$, we require that $\~\psi_1$ be orthogonal to
$\~\psi_0$, hence $\zeta=0$, and
\[
\~\psi_1(z)=0.    \label{e:psi1sol}
\]

Now we determine the solution $\psi_2$ to Eq.~\eqref{e:perteqc}.
Utilizing solutions $\psi_0$ and $\psi_1$ in Eqs.
(\ref{e:envelopeA}) and (\ref{e:psi1sol}), $\psi_2$ can be written
as
\[
\psi_2(x,z) =-[\partial^{-2}_x g(x)]A^3(z)+\~\psi_2(z)\,.  \label{e:316}
\]
When this solution is inserted into  equation
(\ref{e:perteqe}), the solvability condition of this equation gives
\[
\left({\d^2/\d z^2}-\mu+3A^2 \right) \~\psi_2=-3\alpha A^5\,,
\label{e:eqpsi2}
\]
where
\begin{equation}  \label{e:alpha}
\alpha \equiv \langle (\partial^{-1}_x g)^2 \rangle > 0.
\end{equation}
The bounded solution $\~\psi_2$ to this equation is
\[
\~\psi_2(z) = \alpha(A^3-2a^2 A).  \label{e:318}
\]
Combining ~\eqref{e:psi0}, ~\eqref{e:psi1}, ~\eqref{e:psi1sol}, ~\eqref{e:316} and ~\eqref{e:318}, the
perturbation series solution for the solitary wave $\psi(x,z)$ of
Eq.~\eqref{e:NLSEamp} takes the form
\[
\psi(x,z)=A(z)+\epsilon^2 \bigg\{ -\left[\partial^{-2}_x
g(x)\right]A^3(z) +\alpha\left[A^3(z)-2a^2 A(z)\right]\bigg\}+
O(\epsilon^3)\,, \label{e:pertsoln}
\]
where $A(z)$ is given by (\ref{e:envelopeA}).

Notice that the location $z_0$ of the peak of the function $A(z)$ in
Eq.~\eqref{e:pertsoln} is arbitrary at this stage, since  equation
(\ref{e:psi0eq}) which determines $A(z)$ is translation invariant.
However, the original equation (\ref{e:NLSEamp}) for the solitary
wave is not translation invariant due to the presence of the
nonlinear lattice $g(x)$, and it is unlikely that the solitary wave
can be arbitrarily located relative to this nonlinear lattice.
Indeed, a very similar situation arises in a linear lattice
\cite{Yang_SIAM,Hwang_stability,Pelinovsky_2004}, and there it was
shown that solitary waves can only be located at two positions
relative to the lattice. In the present problem, the result turns
out to be similar: only two values of $z_0$ are permissible for
truly localized solitary waves in a nonlinear lattice, as will be
established in the next section by utilizing exponential
asymptotics.

Before ending this section, it should be pointed out that
truly localized solitary waves $\psi(z)$ of Eq. (\ref{e:NLSEamp}) must satisfy
\[  \label{e:constraint}
\int_{-\infty}^{\infty} g'(z/\epsilon) \psi^4(z) dz=0.
\]
This constraint can be obtained by multiplying Eq. (\ref{e:NLSEamp})
with $\psi'(z)$ and then integrating from $-\infty$ to $\infty$. An
analogous constraint was noted
in \cite{Pelinovsky_2004} for the linear lattice problem, and it was pointed out that, if the
lattice is symmetric, this constraint predicts only two
possible locations for truly localized solitary waves
--- one at the point of symmetry and the other half a period
away from it \cite{Yang_SIAM,Pelinovsky_2004}. But for general
asymmetric linear lattices, it does not seem feasible to determine
the locations of solitary waves based on this constraint alone
\cite{Hwang_stability}. Similarly, in the problem at hand, if the
nonlinear lattice $g(z/\epsilon)$ is symmetric,  the  constraint
(\ref{e:constraint}) can also predict the two locations of true
solitary waves; if the lattice is asymmetric, however, this approach
would again fail. The reason is that, when the perturbation series
solution (\ref{e:pertsoln}) is substituted into the constraint
(\ref{e:constraint}), all terms in the  series make contributions of
the same order of magnitude to the integral in (\ref{e:constraint}).
Hence it does not seem possible to solve for $z_0$ without having
obtained all terms in the perturbation series (\ref{e:pertsoln}).
The same difficulty also appears  in the linear lattice problem
\cite{Yang_SIAM,Hwang_stability} and suggests the need for a
perturbation theory beyond all orders. This task is taken up below
by employing an exponential-asymptotics procedure in the wavenumber
domain \cite{Hwang_stability,Akylas_JFM1997}.

%%%%%%%%%%%%%%%%%%%%%%%%%%%%%%%%%%%%%%%%%%%%%%%%%%%%%%%%%%%%%%%%%%%%%%%%%%%%%%%
\section{Growing tails of exponentially small amplitude}
\label{s:expasympt}
%%%%%%%%%%%%%%%%%%%%%%%%%%%%%%%%%%%%%%%%%%%%%%%%%%%%%%%%%%%%%%%%%%%%%%%%%%%%%%%

In this section, we determine the location of the solitary wave
$\psi(z)$ by the exponential asymptotics method. Our approach
closely resembles that for gap solitons in a linear lattice
\cite{Hwang_stability}, thus only the key ideas and steps will be given.
For further details, we refer the reader to \cite{Hwang_stability}
(see also \cite{Akylas_JFM1997}).

The ensuing analysis is based on the fact that, if
 $\psi(z)$  given by
(\ref{e:pertsoln}) is required to decay upstream ($z\ll -1$), then this solution of Eq. (\ref{e:NLSEamp})
 will contain a growing tail  downstream ($z\gg 1$) for generic values
of the position $z_0$ of the solitary wave core. Specifically, if the upstream
asymptotics of $\psi(z)$, as given by the leading-order term in
the perturbation expansion~\eqref{e:pertsoln}, is
\[
\psi\sim 2 a\,\e^{(z-z_0)/\beta}, \qquad z\to -\infty\,,
\label{e:upstream0}
\]
then the downstream asymptotics of the solution will be
\[
\psi\sim 2 a \,\e^{-(z-z_0)/\beta}+H \, \e^{(z-z_0)/\beta}, \qquad
z\gg 1\,, \label{e:downstream0}
\]
where $H(\epsilon, z_0)$ is the growing-tail amplitude. As we
will show later, $H$ vanishes only at two special values of $z_0$
(relative to the nonlinear lattice), thus only two truly-localized
solitary waves exist. The tail amplitude $H(\epsilon, z_0)$ turns
out to be exponentially small in $\epsilon$, so this growing tail
can not be captured by the perturbation series (\ref{e:pertsoln})
and has to be obtained by carrying this expansion in powers of $\epsilon$ beyond all orders.

Following the exponential asymptotics procedure in the wavenumber
domain \cite{Hwang_stability}, we introduce the Fourier transform of
$\psi(x,z)$ with respect to the slow variable $z$,
\[
\widehat \psi(x,K)={1\over 2\pi} \int_{-\infty}^{\infty} \psi(x,z)\e^{-iKz} \d z\,.
\label{e:FT}
\]
Substituting the perturbation series solution~\eqref{e:pertsoln} for
$\psi(x,z)$ into this Fourier transform, we find that
\[
\widehat\psi(x,K)={\sqrt{2}\over 2}\e^{-iKz_0}\sech\bigg({\pi\over
2}\beta K\bigg)\left\{1 +\epsilon^2a^2\,\left[{1\over 2}(1+\beta^2
K^2)\nu(x)-2 \alpha\right]+\cdots\right\}\,,
\label{e:psihatexpansion}
\]
where $\nu(x)=\alpha-\partial^{-2}_x g(x)$. This expansion in the
wavenumber domain is disordered when $K=O(1/\epsilon)$, suggesting
that $\widehat\psi$ has pole singularities at $K=O(1/\epsilon)$. The
residues of these singularities are exponentially small due to the
exponentially small value of the sech function in
(\ref{e:psihatexpansion}) at $K=O(1/\epsilon)$. As we will show
later, these singularities of exponentially-small residue contribute
exponentially-small but growing tails in the physical solution
$\psi(x,z)$. Thus the main goal is to determine the locations and
residues of pole singularities in $\widehat\psi$.

To this end, we replace the disordered expansion (\ref{e:psihatexpansion})
by a uniformly-valid expression,
\[
\widehat\psi= \e^{-iKz_0}\sech\bigg({\pi\over 2}\beta K\bigg)U(x,\kappa)\,,\label{e:eqpsihat}
\]
where $\epsilon K=\kappa$. We then take the Fourier transform of Eq.
(\ref{e:eqmulitscale}) with respect to $z$,
\[
\widehat\psi_{xx}+2i\kappa\widehat\psi_x-\kappa^2\widehat\psi+\epsilon^2(1+g(x))\widehat{\psi^3}
-\epsilon^2\mu\,\widehat\psi=0\ , \label{e:FTeq}
\]
and upon substituting Eq. (\ref{e:eqpsihat}) into \eqref{e:FTeq}, we obtain
the following  equation for $U$
\begin{align}
U_{xx}  & +2i\kappa U_x -\kappa^2 U- \epsilon^2\mu U
\nonumber\\
& \hspace{-0.8cm} +(1+g(x))\cosh{\pi\beta\over 2}K
\int_{-\infty}^{\infty}\d\lambda {U(x,\kappa-\lambda)\over
\cosh{\pi(\kappa-\lambda)\beta\over 2\epsilon}}
\int_{-\infty}^{\infty}\d\rho {U(x,\lambda-\rho) U(x,\rho)\over
\cosh{\pi(\lambda-\rho)\beta\over 2\epsilon} \cosh{\pi\rho\beta\over
2\epsilon}}=0 \label{e:integraleq}\,.
\end{align}

\subsection{The recurrence equation}
In the limit $\epsilon\to 0$ and for $\kappa$ away from
singularities of $U(x, \kappa)$, the main contribution to the double
integral in Eq.~\eqref{e:integraleq} comes from $0<\lambda<\kappa$
when $\kappa
>0$ and from $\kappa<\lambda<0$ when $\kappa<0$. The integral
equation (\ref{e:integraleq}) then simplifies to
\cite{Hwang_stability,Akylas_JFM1997}
\[
U_{xx}   +2i\kappa U_x -\kappa^2 U
+4(1+g(x))\int_{0}^{\kappa\!\!}\d\lambda
U(x,\kappa-\lambda)\int_{0}^{\lambda\!\!}\d\rho
U(x,\lambda-\rho)U(x,\rho) = 0\,. \label{e:approxintegraleq}
\]
The solution to this simplified integral equation can be posed as a
power series in $\kappa$,
\[
U(x,\kappa)={\sqrt{2}\over 2} \sum_{n=0}^{\infty} U_n(x)\kappa^n \,.
\label{e:seriesU}
\]
Substituting this power series into \eqref{e:approxintegraleq}, we
obtain the following recurrence equation for $U_n$,
\begin{align}
{\d^2 U_{n+2}\over \d x^2} = & \ U_n-2i{\d U_{n+1}\over \d x}   \nonumber \\
& -2(1+g(x)) \sum_{m=0}^n U_{n-m}{(n-m)!\over (n+2)!}\sum_{r=0}^m
U_rU_{m-r} r!(m-r)!\,. \label{e:recurrence}
\end{align}
To be consistent with (\ref{e:psihatexpansion}), we set
\[
U_0=1, \quad U_1=0,    \label{e:U0U1}
\]
which serve as the initial conditions for the recurrence iteration
(\ref{e:recurrence}). Note that this recurrence system does not involve
  $\mu$ and $\epsilon$; it only depends on the
functional form of the periodic lattice $g(x)$.

%%%%%%%%%%%%%%%%%%%%%%%%%%%%%%%%%%%%%%%%%%%%%%%%%%%%%%%%%%%%%%%%%%%%%%%%%%%%%%%%%%%%%%%%%%%%%%%%%%%%%%%%%%%%%%%%%%%%%%
\subsection{Behavior near singularities}
%%%%%%%%%%%%%%%%%%%%%%%%%%%%%%%%%%%%%%%%%%%%%%%%%%%%%%%%%%%%%%%%%%%%%%%%%%%%%%%%%%%%%%%%%%%%%%%%%%%%%%%%%%%%%%%%%%%%%%

When $\kappa$ is near the singularities of $U(x,\kappa)$, the
reduced integral equation~\eqref{e:approxintegraleq} does not apply.
We now examine the behavior of $U(x,\kappa)$ near its
singularities, based on the original integral
equation~\eqref{e:integraleq}.

These
singularities occur at $\kappa \approx \kappa_0$, where the linear
part of Eq.~\eqref{e:integraleq} is satisfied, i.e.,
\[
U_{xx}+2i\kappa_0 U_x-\kappa^2_0 U=0\,.
\]
The bounded solution to this linear equation is $U\sim
\e^{-i\kappa_0 x}$. Since $U$ is expected to be $d$-periodic in $x$
in view of (\ref{e:eqpsihat}), therefore, $\kappa_0=\pm 2\pi/d$.
Singularities near $\kappa_0=\pm 4\pi/d, \pm 6\pi/d,\dots$ are also
possible, but they are subdominant and will not be considered.  To
avoid ambiguity, we denote
\[
\kappa_0=2\pi/d\,,
\]
and examine singularities at $\kappa\approx \pm \kappa_0$.

In order to determine the behavior of the solution $U$ near the
singularity $\kappa\approx \kappa_0$, we first introduce the `inner'
wavenumber
\[
\xi={\kappa-\kappa_0\over \epsilon}\,,
\]
that is, $\kappa=\kappa_0+\epsilon\xi$ with $\xi=O(1)$. Guided by
the analysis in~\cite{Hwang_stability,Akylas_JFM1997}, we expand
\[
U={\e^{-i\kappa_0 x}\over \epsilon^4}\big(\Phi_0(\xi)+\epsilon^2
f(x,\xi)+\cdots\big)\,. \label{e:Uexpansion}
\]
The reason for the leading-order term in this expansion being
$O(\epsilon^{-4})$ will be explained later (see
Eq.~\eqref{e:asymptPhi}).

Now we examine the integral equation~\eqref{e:integraleq} near the
singularity $\kappa\approx \kappa_0$. The dominant contributions to
the double integral in \eqref{e:integraleq} come from the following
regions: (i) $\lambda\sim 0$ and $\rho\sim 0$, (ii) $\lambda\sim
\kappa$ with $\rho\sim 0$ and with $\rho\sim \kappa$. Taking into
account the leading-order term near $\kappa\sim\kappa_0$ in
\eqref{e:Uexpansion} and the leading order term near $\kappa\sim 0$
in (\ref{e:seriesU}), we can calculate these dominant contributions,
and Eq.~\eqref{e:integraleq} near $\kappa\sim\kappa_0$ yields
\[
\~U_{xx}+2i\epsilon\xi \~U_x-\epsilon^2\xi^2\~U-\epsilon^2\mu\~U
 +{3(1+g(x))\over \epsilon^2}\int_{-\infty}^{\infty} \omega  \e^{\pi\beta \omega/2}
\csch{\pi \beta \omega \over 2} \Phi_0(\xi-\omega) \hspace{0.04cm}
\d \omega=0\,,\label{e:integraleqUtilde}
\]
where $U=\e^{-i\kappa_0 x}\~U$. Substituting the expansion
~\eqref{e:Uexpansion} into ~\eqref{e:integraleqUtilde}, the terms of
$O(\epsilon^{-4})$ and $O(\epsilon^{-3})$ are automatically
balanced. At $O(\epsilon^{-2})$ we have
\[
\partial^2_x f=\xi^2\Phi_0(\xi)+\mu\Phi_0(\xi)
-3(1+g(x))\int_{-\infty}^{\infty} \omega  \e^{\pi\beta \omega/2}
\csch{\pi \beta \omega\over 2} \Phi_0(\xi-\omega) \hspace{0.04cm} \d
\omega \ \label{e:integraleqUtilde1},
\]
and the solvability condition requires that the $x$-average of the
right-hand side of Eq.~\eqref{e:integraleqUtilde1} equals to zero:
\[
(\mu+\xi^2)\Phi_0(\xi) -3\int_{-\infty}^{\infty} \omega \e^{\pi\beta
\omega/2} \csch{\pi \omega \beta\over 2} \Phi_0(\xi-\omega)
\hspace{0.05cm} \d \omega =0 \,.\label{e:integraleqPhi0}
\]
Recalling that $\beta=1/\sqrt\mu$, this integral equation is identical to that encountered earlier in our analysis of gap solitons in a linear lattice ~\cite{Hwang_stability}. Hence, its solution is
\cite{Hwang_stability,Akylas_JFM1997}
\[
\Phi_0(\xi)={6 \beta^4\over 1+\beta^2\xi^2}\int_{\cal
L^\pm} {1\over \sin^2 \hspace{-0.06cm} s} \phi(s)e^{-s\beta\xi} \d
s, \label{e:integralPhi0}
\]
where
\[
\phi(s)=C\bigg( {2\over \sin s}+{\cos^2 \hspace{-0.07cm} s\over \sin
s}-{3s\cos s\over \sin^2 \hspace{-0.07cm} s}\bigg),
\]
the contours $\cal L^\pm$ extend from $0$ to $\pm i\infty$ for
$\mbox{Im}(\xi)< 0$ and $\mbox{Im}(\xi)> 0$ respectively, and $C$ is
a complex constant which will be determined later. Note that the
function $\Phi_0(\xi)$ given by~\eqref{e:integralPhi0} is analytic
everywhere in the complex plane $\Complex$, save for two simple
poles at $\xi=\pm i/\beta$.  Also, $\Phi_0(\xi)$ satisfies
the integral equation (\ref{e:integraleqPhi0}) only in the complex
$\xi$-plane outside the strip $-1/\beta<\mbox{Im}(\xi)<1/\beta$, as
was explained in~\cite{Hwang_stability}.

The behavior of the function $\Phi_0(\xi)$ near its singularities
$\xi=\pm i/\beta$ is
\[
\Phi_0(\xi)\to -\frac{C \beta^4}{1+\beta^2\xi^2}  \qquad (\xi\to \mp
\frac{i}{\beta}), \label{e:eqPhi0}
\]
and its large-$\xi$ asymptotics is
~\cite{Hwang_stability,Akylas_JFM1997}
\[
\Phi_0(\xi)\to {12C\over 5} {1\over \xi^4} \qquad (\xi\to \infty)\,.
\label{e:asymptPhi}
\]
This fourth-order algebraic decay rate of $\Phi_0(\xi)$ at large $\xi$ implies
the order $\epsilon^{-4}$ in Eq.~\eqref{e:Uexpansion}. From
Eq.~\eqref{e:eqPhi0}, we then obtain  the behavior of
$\widehat\psi(x, K)$ near $K=\kappa_0/\epsilon \mp i/\beta$ as
\[
\widehat\psi\sim {\beta^3 C\over \epsilon^3}
e^{-\pi\beta\kappa_0/2\epsilon}e^{\mp
z_0/\beta}{e^{-i\kappa_0(x+x_0)}\over K-{\kappa_0\over\epsilon} \pm
{i\over\beta}} \qquad (K\to  {\kappa_0 \over\epsilon} \mp {i\over
\beta})\,. \label{e:eqsingularities0}
\]
From the symmetry of the Fourier transform for real functions,
$\widehat\psi$ is also singular at $K=-\kappa_0/\epsilon \mp
i/\beta$, and
\[
\widehat\psi\sim -{\beta^3 C^*\over \epsilon^3}
e^{-\pi\beta\kappa_0/2\epsilon}e^{\mp
z_0/\beta}{e^{i\kappa_0(x+x_0)}\over K+{\kappa_0\over\epsilon} \pm
{i\over\beta}} \qquad (K\to  -{\kappa_0 \over\epsilon} \mp {i\over
\beta})\,. \label{e:eqsingularities}
\]

%%%%%%%%%%%%%%%%%%%%%%%%%%%%%%%%%%%%%%%%%%%%%%%%%%%%%%%%%%%%%%%%%%%%%%%%%%%%%%%%%%%%%%%%%%%%%%%%%%%%%%%%%%%%%%%%%%%%%%
\subsection{The growing tail and locations of solitary waves}
%%%%%%%%%%%%%%%%%%%%%%%%%%%%%%%%%%%%%%%%%%%%%%%%%%%%%%%%%%%%%%%%%%%%%%%%%%%%%%%%%%%%%%%%%%%%%%%%%%%%%%%%%%%%%%%%%%%%%%

We now take the inverse Fourier transform of $\widehat\psi(x, K)$ in order to
calculate the growing tails in $\psi(x,z)$ due to the pole
singularities~\eqref{e:eqsingularities0}--~\eqref{e:eqsingularities}.
The inverse Fourier transform is
\begin{equation}
\label{e:IFTpsi}
\psi(x,z)=\int_{\cal C} \widehat\psi(x,K) e^{iKz}\d K\,,
\end{equation}
where the contour $\cal C$ is taken along the line $\Im(K)=-1/\beta$
and to pass below the poles at $K=\pm\kappa_0/\epsilon -i/\beta$
(the reason for this choice of the contour is explained in
\cite{Hwang_stability}). Also, the contour $\cal C$ passes above the
pole of $\sech(\pi\beta K/2)$ at $K=-i/\beta$, to be consistent with
the desired upstream behavior of $\psi$ in (\ref{e:upstream0}).

Indeed, for $z\ll -1$, the dominant contribution to the inverse Fourier
transform (\ref{e:IFTpsi}) comes from the pole of $\sech(\pi\beta
K/2)$ at $K=-i/\beta$ and one obtains the upstream solution
(\ref{e:upstream0}). On the other hand, for $z\gg 1$, both the
singularities \eqref{e:eqsingularities0}--~\eqref{e:eqsingularities}
at $K=\pm \kappa_0/\epsilon -i/\beta$ and the singularity of
$\sech(\pi\beta K/2)$ at $K=i/\beta$ contribute, and one obtains the
downstream solution behavior as
\[
\psi\sim 2 a \e^{-(z-z_0)/\beta}+{4\pi\beta^3 \widehat C\over
\epsilon^4} \e^{-\pi\beta\kappa_0/2\epsilon}
\sin(\kappa_0x_0-\theta)\e^{(z-z_0)/\beta} \qquad (z\gg
1)\,,\label{e:downstream}
\]
where $\widehat C>0$ and $\theta$ are the amplitude and phase of the
complex constant $C$,
\[
C=\widehat C\e^{i\theta}\,.
\]
Equation~\eqref{e:downstream} is one of the key results in this
paper. It shows that a growing tail of exponentially small amplitude
appears far downstream in the solution $\psi(x,z)$, except when
\[
\sin(\kappa_0x_0-\theta)=0\,,
\]
i.e.,
\[
x_0=\theta/\kappa_0\,,\quad (\theta +\pi)/\kappa_0\,,
\label{e:x0value}
\]
Thus, exactly two solitary waves, located at these values of $x_0$,
are obtained in the nonlinear lattice equation ~\eqref{e:NLSEamp}.

\subsection{Determination of the constant $C$}

It remains to determine the complex constant $C$. This constant
cannot be obtained from the local analysis around the singularities
$\kappa\sim \pm \kappa_0$, but it can be computed by solving the
recurrence relation (\ref{e:recurrence}). For this purpose, we
derive the asymptotics of the recurrence functions $U_n$ for large
$n$, which depends on $C$. The derivation is based on the idea that
the `inner' solution (\ref{e:Uexpansion}) of $U(x,\kappa)$ near the
singularities, when $\xi\to \infty$, must match the `outer'
solution (\ref{e:seriesU}) of $U(x,\kappa)$ away from the
singularities. First, from the inner expansion (\ref{e:Uexpansion})
and the asymptotics (\ref{e:asymptPhi}), we see that
\[
U(x, \kappa) \sim \frac{12C}{5} {1\over (\kappa-\kappa_0)^4}
e^{-i\kappa_0 x}  \qquad (\epsilon \ll |\kappa-\kappa_0| \ll 1),
\label{e:expansionUtilde}
\]
or
\[
U\sim \frac{12\widehat{C}}{5} {1\over
(\kappa-\kappa_0)^4}\left[\cos(\kappa_0x-\theta)-i\sin(\kappa_0x-\theta)\right]
\qquad (\epsilon \ll |\kappa-\kappa_0| \ll 1). \label{e:expansionUa}
\]
From the symmetry relation $U(x,\kappa)=U^*(x,-\kappa*)$ for real
functions $\psi(x,z)$, we also have
\[
U\sim \frac{12\widehat{C}}{5} {1\over
(\kappa+\kappa_0)^4}\left[\cos(\kappa_0x-\theta)+i\sin(\kappa_0x-\theta)\right]
\qquad (\epsilon \ll |\kappa+\kappa_0| \ll 1). \label{e:expansionUb}
\]
These two asymptotic expressions can be combined and re-written as
\[
U\sim {192\kappa_0^4 \widehat{C} \over 5(\kappa^2-\kappa_0^2)^4}
\left[ \cos(\kappa_0x-\theta)-i {\kappa\over \kappa_0}\sin
(\kappa_0x-\theta)\right] \qquad (\epsilon \ll |\kappa\pm \kappa_0|
\ll 1). \label{e:expansionUc}
\]
This expression is consistent with
(\ref{e:expansionUa})--(\ref{e:expansionUb}) near the singularities.
More importantly, it has the  property that, when expanded
into power series of $\kappa$, the coefficients of all even powers of
$\kappa$ are purely real and the coefficients of all odd powers of
$\kappa$ are purely imaginary, as required by  the
outer solution (\ref{e:seriesU}), with $U_n$ given by the recurrence
relation (\ref{e:recurrence}) under the initial conditions
(\ref{e:U0U1}).

Expanding (\ref{e:expansionUc}) into power series of $\kappa$ and
 requiring this expansion to be consistent with the outer solution
(\ref{e:seriesU}), we obtain the asymptotic behavior of $U_n$ for
$n\gg 1$ as
\[
U_{2m}\sim \mathbb D {m^3\over \kappa_0^{2m}} \cos (\kappa_0
x-\theta) \,, \qquad  U_{2m+1}\sim -i \,\mathbb D {m^3\over
\kappa_0^{2m+1}} \sin (\kappa_0 x-\theta)  \,,
\label{e:asymptoticUm}
\]
where the coefficient $\mathbb D$ is related to $\widehat{C}$ by
\[
\label{e:Chatformula} \widehat{C}=\frac{5\sqrt{2}}{64}\kappa_0^4
\,\mathbb D\,.
\]
Thus, by solving the recurrence relation~\eqref{e:recurrence} and
from its large-$n$ asymptotics, we can obtain $\mathbb D$ and
$\theta$, and hence the complex constant $C$. Since the
recurrence equation~\eqref{e:recurrence} depends only on the lattice
function $g(x)$, the constant $C$  also depends only on the
lattice $g(x)$ and not on $\mu$, $\epsilon$.

%%%%%%%%%%%%%%%%%%%%%%%%%%%%%%%%%%%%%%%%%%%%%%%%%%%%%%%%%%%%%%%%%%%%%%%%%%%%%%%%%%%%%%%%%%%%%%%%%%%%%%%%%%%%%%%%%%%%%%
\section{Linear stability problem}
\label{s:stability}
%%%%%%%%%%%%%%%%%%%%%%%%%%%%%%%%%%%%%%%%%%%%%%%%%%%%%%%%%%%%%%%%%%%%%%%%%%%%%%%%%%%%%%%%%%%%%%%%%%%%%%%%%%%%%%%%%%%%%%

In this section, we determine the linear stability of the two
solitary waves whose locations are given by Eq.~\eqref{e:x0value}.
We will show that the solitary wave located at $x_0=\theta/\kappa_0$
is linearly stable, while that located at $(\theta +\pi)/\kappa_0$ is
linearly unstable, and the unstable eigenvalue is exponentially
small in $\epsilon$. This calculation follows the approach used in
\cite{Hwang_stability,Calvo_stability} for the stability analysis of
solitary wavepackets of the fifth-order KdV equation and gap
solitons of the NLS equation with a linear lattice.

Let $\psi_s(z)$ be a solitary wave of Eq.~\eqref{e:NLSEamp}, whose
leading-order term $\psi_0=A(z)$ in (\ref{e:envelopeA}) is centered
at $z_0=z_{0s}$, where $z_{0s}=\epsilon x_{0s}$, and $x_{0s}$ is one
of the two positions given in Eq.~\eqref{e:x0value}. Consider the
perturbed solution
\[
\Psi(z, t)=e^{-i\mu t}\left\{\psi_s(z)+[v(z)+w(z)]\e^{\lambda
t}+[v^*(z)-w^*(z)]\e^{\lambda^*t}\right\},
\]
where $v, w\ll 1$ are normal-mode perturbations. Substituting this
perturbed solution into Eq. (\ref{e:NLSE}) and neglecting nonlinear
terms in $(v,w)$, we obtain the linear-stability eigenvalue problem
\cite{Yang_SIAM}
\[
L_0L_1v=-\lambda^2 v.  \label{e:eigenvalproblem}
\]
where
\[
L_0 ={\d^2\over \d z^2}+(1+g(x))\psi_s^2 -\mu, \qquad L_1={\d^2\over
\d z^2}+3(1+g(x))\psi_s^2 -\mu. \label{e:definitionL1}
\]
If there exist eigenvalues $\lambda$ with positive real parts, then
the solitary wave is linearly unstable. Otherwise it is linearly
stable.

When $\epsilon\ll 1$, the discrete eigenvalue $\lambda$ is small. To
calculate this eigenvalue, we expand $v$ into powers of $\lambda$,
\[
\label{e:vexpansion}
v=v_0 +\lambda^2 v_1+\lambda^4 v_2+\cdots.
\]
Substituting this expansion into Eq.~\eqref{e:eigenvalproblem}, at $O(1)$, we have
\[
L_0L_1v_0=0.     \label{e:L0L10}
\]
Recall that
\[
L_0\psi(x,z; x_0)=0\,,   \label{e:L0psi}
\]
where $\psi(x,z; x_0)$ is the nonlocal solution to Eq.
(\ref{e:NLSEamp}) whose leading term $\psi_0=A(z)$ in
(\ref{e:envelopeA}) is centered at $z=\epsilon x_0$. Taking the
derivative of this equation with respect to $x_0$ and then setting
$x_0=x_{0s}$, we get
\[
L_1 \left. \frac{\partial\psi}{\partial x_0} \right|_{x_0=x_{0s}}=0\,.
\]
Hence the solution to Eq.~\eqref{e:L0L10} is
\[
\label{e:v0form}
v_0=\left. \frac{\partial\psi}{\partial x_0} \right|_{x_0=x_{0s}}.
\]
Notice from Eq.~\eqref{e:downstream} that
\[
\label{e:v0_tail} v_0 \sim \frac{2\epsilon a}{\beta}
\e^{-(z-z_{0s})/\beta}+{4\kappa_0\pi\beta^3 \widehat{C}\over
\epsilon^4}
\e^{-\pi\beta\kappa_0/2\epsilon}\cos(\kappa_0x_{0s}-\theta)
\e^{(z-z_{0s})/\beta}, \quad  z \gg 1,
\]
thus $v_0$ contains a growing tail and is nonlocal. Since the discrete
eigenfunction $v$ must be localized,  this growing tail in $v_0$
must be cancelled by another growing tail in $\lambda^2 v_1$.

The equation for $v_1$ is found
from Eq.~\eqref{e:eigenvalproblem} at $O(\lambda^2)$ as
\[
L_0 L_1v_1=-v_0.   \label{e:L0L1v1}
\]
Letting $L_1v_1=w_0$, we first solve
\[
\label{e:w0eqn}
L_0w_0=-v_0.
\]
Since the growing tail in $v_0$ is exponentially small, its
contribution to a likewise exponentially-small growing tail in $v_1$
(through Eq. (\ref{e:L0L1v1})) can be ignored, because the localized
(algebraically small) terms in $v_0$ will turn out to create a relatively
larger (i.e.,  algebraically small) growing tail in  $v_1$. Thus, in the calculation of $w_0$, it suffices to take
$v_0$ as (\ref{e:v0form}), with $\psi$ given by the perturbation series
(\ref{e:pertsoln}). The corresponding solution $w_0$ can be expanded
into a perturbation series
\[
\label{e:w0expansion} w_0 = \epsilon B(z)+\epsilon^3
\hat{w}_0(x,z)+\cdots.
\]
Inserting this expansion into \eqref{e:w0eqn}, at $O(\epsilon)$ we
obtain
\[
-\partial^2_x \, \hat{w}_0=B''(z)-\mu
B(z)+(1+g(x))A^2(z)B(z)-A'(z)\,.
\]
The solvability condition of this equation yields the governing
equation for $B(z)$ as
\[
{\d^2 B\over \d z^2}-\mu B+A^2 B=A'(z)\,,
\]
hence
\[
B(z)={1\over 2}(z-z_{0s})A(z)\,. \label{e:eqB}
\]

Now we solve for $v_1$ from $L_1v_1=w_0$. This solution can be
expanded as
\[
\label{e:v1expansion} v_1 = \epsilon F(z)+\epsilon^3
\hat{v}_1(x,z)+\dots\,.
\]
Substituting this expansion into $L_1v_1=w_0$, at $O(\epsilon)$ we
get
\[
-\partial^2_x \hat{v}_1=F''(z)-\mu F(z)+3(1+g(x))A^2(z)F(z) -\frac12
(z-z_{0s})A(z)\,.
\]
The solvability condition of this equation gives
\[
{\d^2 F\over \d z^2}-\mu F+3A^2 F=\frac12 (z-z_{0s})A\,.
\label{e:eqF}
\]
Note that the homogeneous solution $A'(z)$ of (\ref{e:eqF}) is not
orthogonal to the inhomogeneous term, hence the solution $F(z)$ to
Eq.~\eqref{e:eqF} is nonlocal. For our purpose, we seek a solution
$F(z)$ so that $F(z)\to 0$ as $z\to -\infty$ and $F(z) \to F_0
\e^{(z-z_{0s})/\beta}$ for $z\gg 1$, where $F_0$ is a constant. To
determine $F_0$, we multiply Eq. (\ref{e:eqF}) by $A'(z)$ and then
integrate from $-\infty$ to $z\gg 1$,
\[
\int_{-\infty}^z A'(\~z) \bigg[F''(\~z)-\mu
F(\~z) +3 A^2(\~z)F(\~z)\bigg] \d \~z ={1\over 2}\int_{-\infty}^z (\~z-
z_{0s})A(\~z) A'(\~z)\,  \d \~z\,.
\label{e:integraleqF}
\]
Employing integration by parts on the left-hand side and using the
large-$z$ asymptotics of $F(z)$ above, we can obtain the left
integral in terms of $F_0$. The right integral approaches a constant
as $z\to +\infty$, which can be readily obtained using the
functional form (\ref{e:envelopeA}) of $A(z)$. After these
calculations, Eq. (\ref{e:integraleqF}) yields $F_0=a\beta^3/8$.
Hence, the corresponding large-$z$ asymptotics of $v_1$ is
\[
v_1 \sim {a\beta^3 \over 8}\epsilon \hspace{0.06cm}
\e^{(z-z_{0s})/\beta} \qquad (z\gg 1)\,.
\]
Inserting this growing tail of $v_1$ and the growing tail of $v_0$
in (\ref{e:v0_tail}) into the expansion \eqref{e:vexpansion} of $v$
and utilizing the relation~\eqref{e:Chatformula}, the eigenvalue
formula is then found to be
\[
\lambda^2=-\frac52\kappa_0^5\pi \beta{\mathbb D}  \cdot
{e^{-\pi\beta\kappa_0/2\epsilon}\over
\epsilon^5}\cos(\kappa_0x_{0s}-\theta)\,,\label{e:formulaeigen}
\]
where $\mathbb D$ and $\theta$ are obtained by solving the
recurrence equation ~\eqref{e:recurrence}. Note that this eigenvalue is exponentially
small in $\epsilon$. In addition, since $\mathbb D >0$, the solitary
wave located at $x_{0s}=\theta/\kappa_0$ is linearly stable, and the
one located at $x_{0s}=(\theta+\pi)/\kappa_0$ is linearly unstable.

We remark in passing that the eigenvalue $\lambda$ given by formula
(\ref{e:formulaeigen}) is two orders larger (in $\epsilon$) than the
eigenvalues found in earlier studies for solitary wavepackets of the
fifth-order KdV equation \cite{Calvo_stability} and Bloch-wavepacket
solitons of the NLS equation with a linear lattice
\cite{Hwang_stability}. However, this does not imply that the linear
instability (for one of the two solitons) in the present nonlinear
lattice problem is stronger than those in the fifth-order KdV
equation and the NLS equation with a linear lattice. The reason is
that  the small parameter $\epsilon$ in
\cite{Hwang_stability,Calvo_stability} measures the wave peak
amplitude, while in the present analysis $\epsilon$ is a long-wave
parameter. Indeed, the peak amplitude of solitary waves in our
analysis is $O(1)$ rather than $O(\epsilon)$ (see Eq.
(\ref{e:psilimit})). In the end of Sec. \ref{sec:preliminaries}, we
mentioned an alternative treatment where, through rescaling, the
solitary wave becomes long and also features low amplitude of
$O(\epsilon)$. This equivalent analysis is the proper counterpart of
those in \cite{Hwang_stability,Calvo_stability}. After the rescaling
(\ref{e:2-8}), in fact, the stability eigenvalue is of the same
order in $\epsilon$ as  in \cite{Hwang_stability,Calvo_stability}.

%%%%%%%%%%%%%%%%%%%%%%%%%%%%%%%%%%%%%%%%%%%%%%%%%%%%%%%%%%%%%%%%%%%%%%%%%%%%%%%%%%%%%%%%%%%%%%%%%%%%%%%%%%%%%%%%%%%%%%
\section{Numerical results}
%%%%%%%%%%%%%%%%%%%%%%%%%%%%%%%%%%%%%%%%%%%%%%%%%%%%%%%%%%%%%%%%%%%%%%%%%%%%%%%%%%%%%%%%%%%%%%%%%%%%%%%%%%%%%%%%%%%%%%

\begin{figure}[h!]
\centerline{
\includegraphics[width=0.7\textwidth]{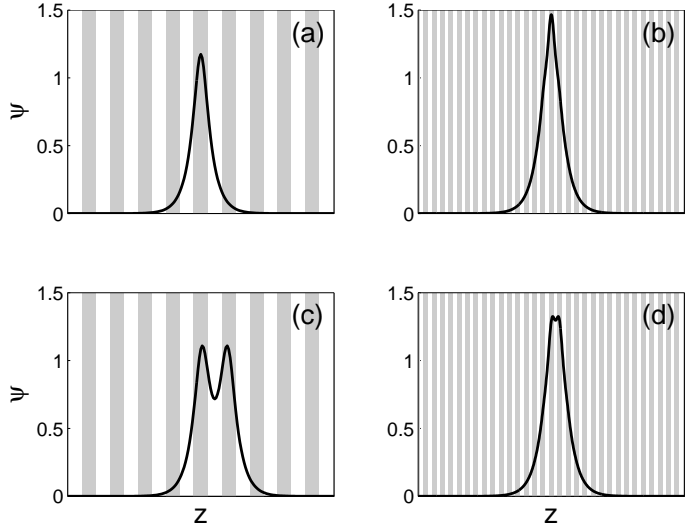}}
\caption{Profiles of on-site and off-site solitary waves in the
nonlinear lattice equation (\ref{e:NLSEamp}) at two values of
$\epsilon=0.5$ and $0.15$ (with $\mu=1$). Upper row: on-site
solutions; lower row: off-site solutions; left column:
$\epsilon=0.5$; right column: $\epsilon=0.15$. The vertical stripes
mark locations of high nonlinear-lattice values. } \label{f:profile}
\end{figure}

In this section, we present numerical results for solitary-wave
profiles and their linear-stability eigenvalues, and make a
comparison with the above analytical results. The numerical
algorithms for these computations can be found in \cite{Yang_SIAM}.
In our computations, we take $\mu=1$ and the nonlinear periodic
lattice to be
\[
\label{Vform}
g(x)=g_0 \cos x\,,
\]
with $g_0=1$. In this case, $d=2\pi$ and $\kappa_0=2\pi/d=1$. The
numerical procedure for solving the recurrence relation
(\ref{e:recurrence}) is similar to that in \cite{Hwang_stability}.
Our computation confirms that $U_n$ for $n\gg 1$ indeed approaches
the asymptotic form (\ref{e:asymptoticUm}) with
\[
\mathbb D = 0.1638\,, \qquad \theta=0\,.
\]
Thus our theory predicts that the two solitary waves are located at
$x_{0}=0$ (maximum of $g(x)$) and $x_0=\pi$ (minimum of $g(x)$), or
equivalently $z_{0}=0$ and $z_0=\epsilon\pi$ respectively. To borrow
the terminology of gap solitons in linear lattices, we call the
solitary wave located at the lattice-maximum $z_{0}=0$ `on-site',
and the other one located at the lattice-minimum $z_0=\epsilon\pi$
`off-site'. Numerically, we have computed these two solitary waves
(for each value of $\epsilon$), and found them to be indeed located
at the two $z_0$ positions. To demonstrate, these solitary waves for
$\epsilon=0.5$ and $0.15$ are displayed in Fig. \ref{f:profile}.
Notice that the on-site solitary waves have a single hump, while the
off-site ones have double humps. In addition, when $\epsilon$ is
small, both on-site and off-site solitary waves have approximately a
sech profile, in agreement with the perturbation series solution
(\ref{e:pertsoln}).

\begin{figure}[h!]
\centerline{\includegraphics[width=0.7\textwidth]{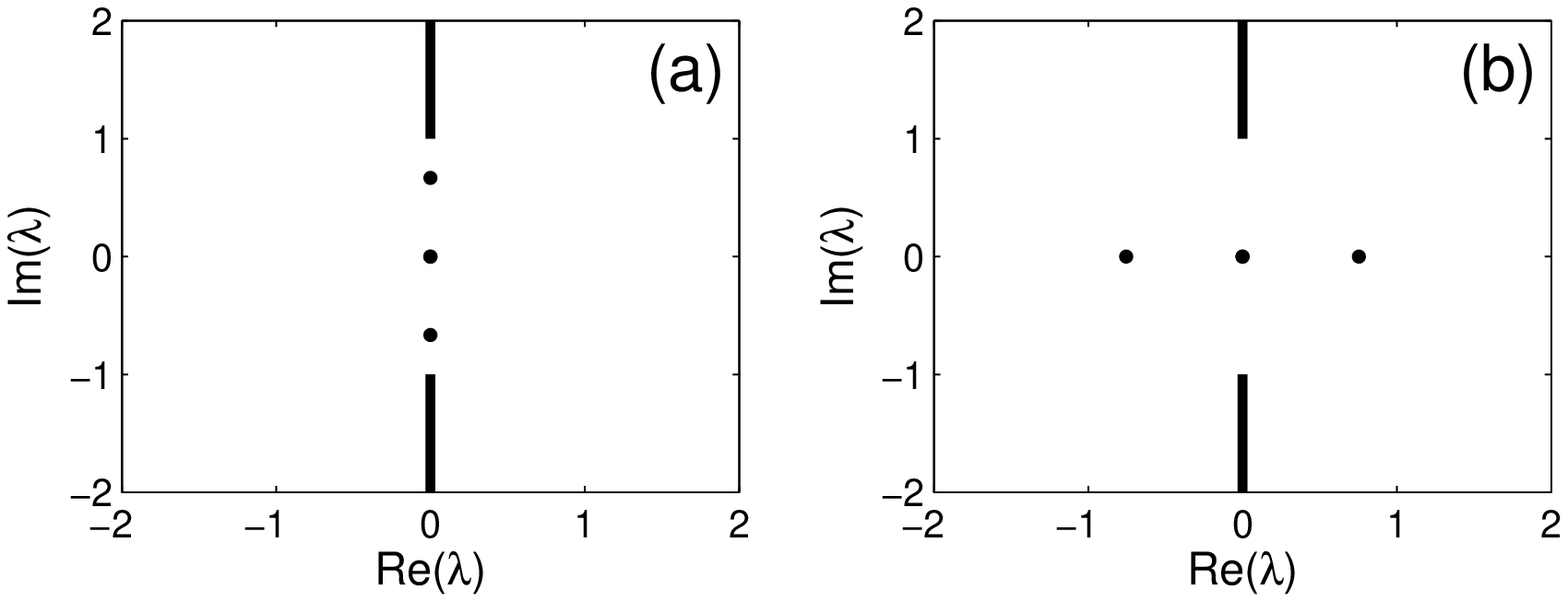}}

\vspace{0.4cm}
\centerline{\includegraphics[width=0.74\textwidth]{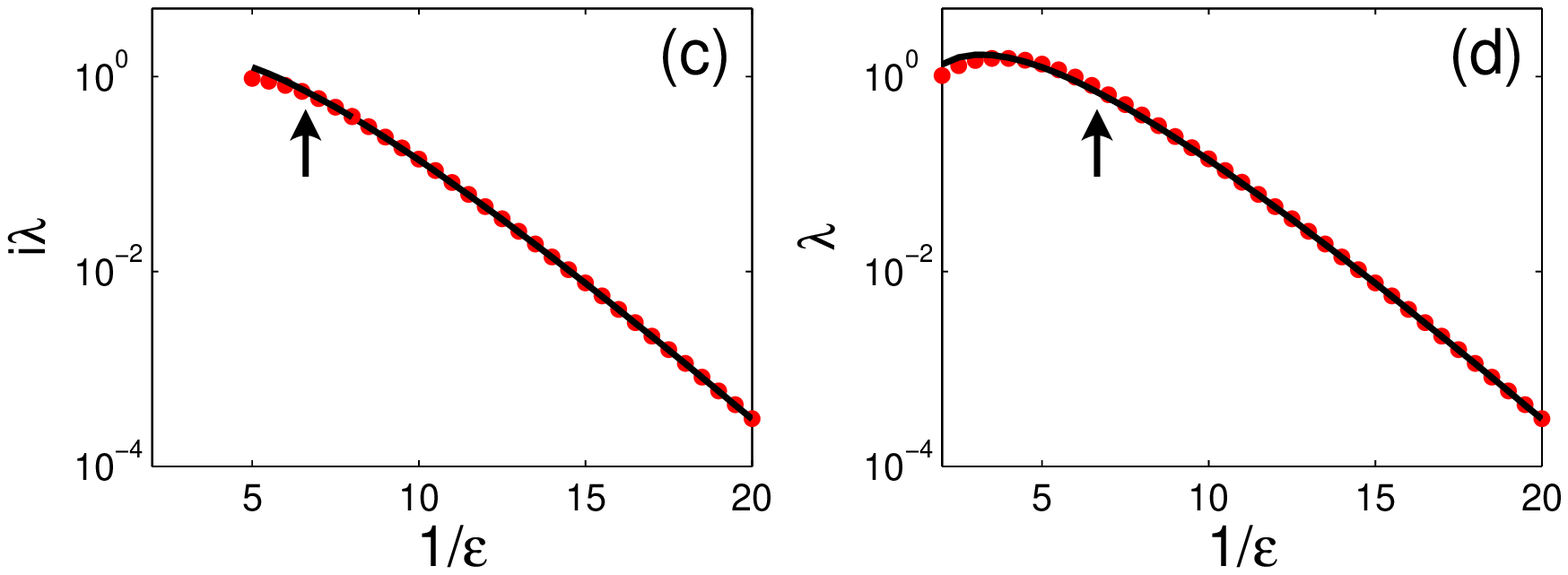}}
\caption{(a,b) Stability spectra for the on-site and off-site
solitary waves in Fig. \ref{f:profile}(b,d) respectively
($\epsilon=0.15$); (c,d) comparison between numerical and analytical
discrete eigenvalues of on-site and off-site solitary waves at
various values of $\epsilon$; dotted lines: numerical values; solid
lines: analytical values from formulae~\eqref{e:formulaeigen}. The
arrows mark locations of the  value of $\epsilon$ in (a,b). }
\label{f:eigenvalue}
\end{figure}

Next we numerically determine the linear stability of these solitary
waves and make a comparison with our analytical results. The whole
linear-stability spectra for the on-site and off-site solitary waves
in Fig. \ref{f:profile}(b,d) at $\epsilon=0.15$ are shown in Fig.
\ref{f:eigenvalue}(a,b). The spectrum in Fig. \ref{f:eigenvalue}(a)
lies entirely on the imaginary axis, indicating that this on-site
solitary wave is linearly stable. In this spectrum, a pair of purely
imaginary discrete eigenvalues can be seen. These are the
counterparts of our analytical imaginary eigenvalues given by Eq.
(\ref{e:formulaeigen}) with $x_{0s}=\theta/\kappa_0=0$. The spectrum
in Fig. \ref{f:eigenvalue}(b) contains a real positive eigenvalue,
indicating that the underlying off-site solitary wave is linearly
unstable, in agreement with our analytical prediction. In
particular, this positive eigenvalue is the counterpart of our
analytical positive eigenvalue given by Eq. (\ref{e:formulaeigen})
with $x_{0s}=(\theta+\pi)/\kappa_0=\pi$.

Now we quantitatively compare the numerical linear-stability
eigenvalues with the analytical formula \eqref{e:formulaeigen}. For
this purpose, we have numerically obtained the discrete eigenvalues
(as those in Fig. \ref{f:eigenvalue}(a,b)) for on-site and off-site
solitary waves at various values of $\epsilon$, and the results are
shown in Fig. \ref{f:eigenvalue}(c,d) by dotted lines. The
analytical eigenvalues from the formulae \eqref{e:formulaeigen} for
the on-site and off-site cases are also plotted as solid lines,
and excellent agreement can be seen for both cases. This
verifies that the analytical formula~\eqref{e:formulaeigen} is
asymptotically accurate.

\begin{figure}[h!]
\centerline{
\includegraphics[width=0.7\textwidth]{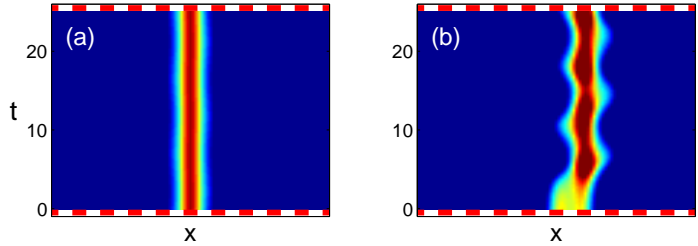}}
\caption{Nonlinear evolutions of the on-site (a) and off-site (b)
solitary waves in Fig. \ref{f:profile}(b,d) under phase-gradient
perturbations (\ref{e:ic}). Red bars represent locations of high
nonlinear-lattice values. } \label{f:evolution}
\end{figure}

Finally, we numerically examine how these solitary waves
evolve nonlinearly under weak perturbations. For this purpose, we again consider
the on-site and off-site solitary waves in Fig. \ref{f:profile}(b,d)
at $\epsilon=0.15$. We perturb these waves initially by a small
phase gradient as
\[
\Psi(z,0)=\psi(z)e^{i\gamma z},   \label{e:ic}
\]
where $\psi(z)$ is the solitary wave and $\gamma=0.01$. This phase-gradient perturbation gives the
solitary wave a small initial `push'. Evolutions of the on-site and
off-site solitary waves under this perturbation are obtained by
simulating Eq. (\ref{e:NLSE}) with the above initial condition
(\ref{e:ic}), and the results are displayed in Fig.
\ref{f:evolution}. It is seen that the on-site solitary wave is not
affected by this perturbation and stays at its initial on-site
position (see Fig. \ref{f:evolution}(a)),  consistent
with the linear stability of this on-site solitary wave
established in Fig. \ref{f:eigenvalue}(a). On the other hand, under
the same perturbation, the off-site solitary wave moves from its
initial off-site position to a nearby on-site position and then
oscillates around it (see Fig. \ref{f:evolution}(b)). When $t\to
\infty$, the oscillation eventually dies out, and the solution
evolves into a stationary on-site solitary wave. This nonlinear
evolution scenario is consistent with the linear instability of this
off-site solitary wave, as shown in Fig. \ref{f:eigenvalue}(b).

\section{Bound states}
In addition to the two fundamental solitary waves obtained earlier,
the nonlinear-lattice equation (\ref{e:NLSEamp}) also admits
higher-order solitary waves, or bound states
\cite{Malomed_nonlinear_lattice}. These can be analytically
constructed by matching the tails of more than one of the nonlocal
solitary waves discussed in Sec. \ref{s:expasympt}. A similar
construction has been detailed in \cite{Akylas_bound_states} for
bound states of the NLS equation with a linear sinusoidal periodic
potential (see also \cite{Akylas_JFM1997}). Here we shall  sketch
the analysis for bound states involving two nonlocal solitary waves
in a nonlinear lattice.

\begin{figure}[h!]
\centerline{
\includegraphics[width=0.7\textwidth]{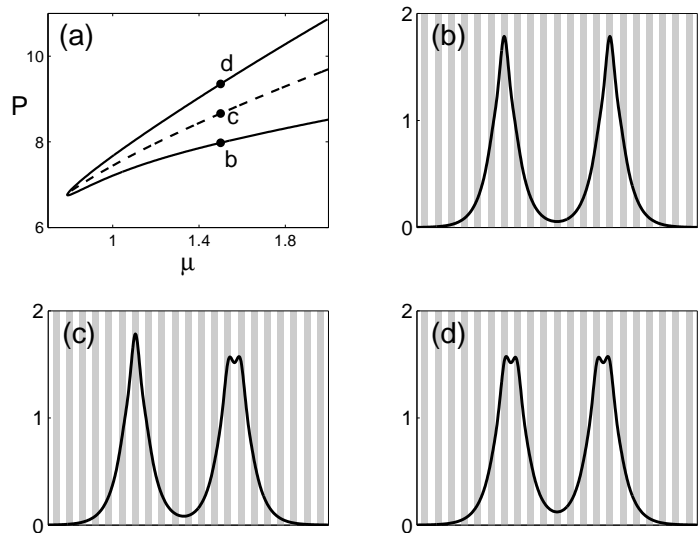}}
\caption{A family of bound states with same polarity in Eq.
(\ref{e:NLSEamp}) for the cosine nonlinear potential (\ref{Vform}) with $g_0=1$
and $\epsilon=0.15$: (a) the power curve; (b,c,d) bound-state
profiles at points of the power branches marked by the same letters
in (a).} \label{f:bound_state}
\end{figure}

For simplicity, we assume the symmetric cosine nonlinear lattice
(\ref{Vform}), with given depth $g_0$.  In this
case, $\kappa_0=1$ and $\theta=0$. Then we consider two nonlocal
solitary waves of (see Sec. \ref{s:expasympt}), $\psi^+(z)$ and
$\psi^-(z)$, whose main `sech' humps are centered at $z_0^+=\epsilon
x_0^+$ and $z_0^-=-\epsilon x_0^-$, respectively, with $z_0^\pm >0$,
and $z_0^+ +z_0^-\gg 1$ (large separation). In addition, let
$\psi^\pm(z)\to 0$ as $z\to\pm \infty$, so that the right-hand tail of
$\psi^-(z)$ and the left-hand tail of $\psi^+(z)$ are nonlocal.
According to (\ref{e:downstream}), the right-hand tail of
$\psi^-(z)$ is given by
\[
\psi^-(z) \sim 2 a \e^{-(z+z_0^-)/\beta}-{4\pi\beta^3 \widehat
C\over \epsilon^4} \e^{-\pi\beta/2\epsilon} \sin x_0^- \
\e^{(z+z_0^-)/\beta} \qquad (z\gg -z_0^-)\,.\label{e:downstreamb}
\]
For symmetric lattice functions $g(x)$, Eq. (\ref{e:NLSEamp}) is
invariant with respect to reflection in $z$ ($z\to -z$). In
addition, if $\psi(z)$ is a solution to (\ref{e:NLSEamp}), so is
$-\psi(z)$. Thus the left-hand tail of $\psi^+(z)$ can be obtained
from (\ref{e:downstream}) after reflection in $z$ as
\[
\psi^+(z) \sim \pm 2 a \e^{(z-z_0^+)/\beta}\mp {4\pi\beta^3 \widehat
C\over \epsilon^4} \e^{-\pi\beta/2\epsilon} \sin x_0^+ \
\e^{-(z-z_0^+)/\beta} \qquad (z\ll z_0^+)\,.\label{e:downstreamc}
\]
Here, the upper sign in (\ref{e:downstreamc}) corresponds to the
case where the main humps of the two nonlocal waves have the
same polarity (sign), while the lower sign pertains to the case of
opposite polarity. To obtain a solitary wave (bound state) comprising
these two nonlocal waves, we require that the right-hand tail
(\ref{e:downstreamb}) of $\psi^-(z)$ and the left-hand tail
(\ref{e:downstreamc}) of $\psi^+(z)$ match smoothly in the
overlap region  $-z_0^-\ll z\ll z_0^+$. This requirement gives
\[
\sin x_0^-=\sin x_0^+= \mp {a \over 2\pi \widehat C}{\epsilon^4\over
\beta^3}
\e^{\pi\beta/2\epsilon}\,\e^{-\epsilon(x_0^++x_0^-)/\beta}\,.
\label{e:matchingcond}
\]
\indent{These matching equations are identical to those in
\cite{Akylas_bound_states} after a scaling in $(x_0^-, x_0^+,
\epsilon, \widehat C)$, and their solutions can be taken directly
from \cite{Akylas_bound_states}. Specifically, for fixed $\mu>0$,
$\epsilon>0$ and given sign (polarity), these equations admit an
infinite number of solutions $(x_0^-, x_0^+)$. Each solution
corresponds to a bound state whose leading-order
approximation is}
\[
\psi(z) \sim a\sech{z+z_0^-\over \beta}\pm a\sech{z-z_0^+\over
\beta},   \label{e:anal_sol}
\]
and a continuous family of bound states is obtained when
$\epsilon$ or $\mu$ varies. Each bound-state family contains triple
branches and, for fixed $\epsilon$, these branches disappear when
$\mu$ falls below a certain threshold (or equivalently, for fixed
$\mu$, these branches disappear when $\epsilon$ falls below a
certain threshold). To demonstrate, we take the cosine nonlinear
potential (\ref{Vform}) with $g_0=1$ and $\epsilon=0.15$. In this nonlinear
lattice, a family of bound states comprising two fundamental
solitons of the same polarity is numerically obtained and displayed
in Fig. \ref{f:bound_state}. The power curve $P(\mu)$ of this
family, defined as
\begin{equation}
P(\mu)=\int_{-\infty}^\infty \psi^2(z; \mu) \hspace{0.07cm} dz,
\end{equation}
contains three branches. On the lower branch, the bound state
comprises two on-site fundamental solitons which are separated
approximately by 8 lattice sites. On the upper branch, the bound
state comprises two off-site fundamental solitons which are
separated approximately by 7 lattice sites. On the middle branch,
the bound state comprises an on-site and an off-site fundamental
soliton which are separated approximately by 7.5 lattice sites.
These solution branches exist only when $\mu>\mu_c\approx 0.7887$;
thus, they do not bifurcate from infinitesimal linear waves at the edge
$\mu=0$ of the continuous spectrum of Eq. (\ref{e:NLSE}). These
numerical results are in good agreement with the theoretical
predictions based on the formulae
(\ref{e:matchingcond})--(\ref{e:anal_sol}).

\section{Concluding remarks}

In this article, we developed an asymptotic theory for long solitary waves and
their linear stability  in a general nonlinear lattice. Based on exponential asymptotics, we showed that  long
solitary waves can only be located at two positions relative to the
nonlinear lattice, regardless of the number of local maxima and
minima in the lattice. In general, these positions are determined by a
certain recurrence relation that includes information beyond all orders of the usual multiple-scale perturbation expansion.  From the same recurrence relation, one may also deduce that, of these two
solitary waves, one is linearly stable and the other is unstable. If the lattice is
symmetric, then the solitary-wave positions are simply the point of symmetry and
half a lattice-period away from it.
 In particular, for
the special cosine lattice, the solitary wave centered at the
maximum/minimum of the lattice is linearly stable/unstable. We also
derived an analytical formula for the linear-stability eigenvalues,
which are exponentially small with respect to the long-wave
parameter (the ratio between the lattice period and the width of the
solitary wave). The predictions of this analytical formula were
compared against numerical results and excellent agreement was
observed. Finally, it was pointed out that an infinite number of
multi-solitary-wave bound states are possible in a nonlinear
lattice, and their analytical construction was presented.

The exponential asymptotics procedure used in this investigation
closely resembles that in \cite{Hwang_stability,Akylas_bound_states}
for linear lattices and in \cite{Akylas_JFM1997,Calvo_stability} for
the fifth-order KdV equation. In fact, the integral equation
(\ref{e:integraleqPhi0}), which plays a key role in the analysis,  actually arises in all these three different
physical models. This suggests that our asymptotic
procedure in the wavenumber domain is a possibly universal treatment of multiscale solitary-wave problems, and it is likely to
find  applications in other physical settings as well.

\section*{Acknowledgment}
The work of G.H. and J.Y. is supported in part by the Air Force
Office of Scientific Research (Grant USAF 9550-09-1-0228) and the
National Science Foundation (Grant DMS-0908167), and the work of
T.R.A. is supported in part by the National Science Foundation
(Grant DMS-098122).

\bigskip
%%%%%%%%%%%%%%%%%%%%%%%%%%%%%%%%%%%%%%%%%%%%%%%%%%%%%%%%%%%%%%%%%%%%%%%%%%%%%%%%%%%%%%%%%%%%%%%%%%%%%%%%%

\begin{center}
UNIVERSITY OF VERMONT \\
MASSACHUSETTS INSTITUTE OF TECHNOLOGY \\
UNIVERSITY OF VERMONT

(Received June 24, 2011)

\end{center}

\end{document}